\documentstyle[12pt,aaspp4]{article}
\accepted{}
\lefthead{Eskridge etal}
\righthead{UV-Optical Pixel Maps}

\received{2002 September 13}
\begin{document}

\title{UV-Optical Pixel Maps of Face-On Spiral Galaxies -- Clues for Dynamics
and Star Formation Histories\footnotemark}

\author{Paul B.~Eskridge\altaffilmark{2}, Jay A.~Frogel\altaffilmark{3,4,5}, 
Violet A.~Taylor\altaffilmark{6}, Rogier A.~Windhorst\altaffilmark{6}, Stephen 
C.~Odewahn\altaffilmark{6}, Claudia A.T.C.~Chiarenza\altaffilmark{6}, 
Christopher J.~Conselice\altaffilmark{7}, Richard de Grijs\altaffilmark{8}, 
Lynn D.~Matthews\altaffilmark{9}, Robert W.~O'Connell\altaffilmark{10}, \& John 
S.~Gallagher, III\altaffilmark{11}}

\footnotetext{Based on observations with the NASA/ESA Hubble Space Telescope,
obtained at the Space Telescope Sciences Institute, which is operated by the
Association of Universities for Research in Astronomy, Inc.~under NASA contract
No.~NAS5-26555.}
\altaffiltext{2}{Department of Physics and Astronomy, Minnesota State
University, Mankato, MN 56001}
\altaffiltext{3}{Department of  Astronomy, The Ohio State University, Columbus,
OH 43210}
\altaffiltext{4}{NASA Headquarters, Code S, 300 E Street SW, Washington, DC 
20546}
\altaffiltext{5}{Visiting Investigator, Department of Terrestrial Magnetism, 
Carnegie Institution of Washington}
\altaffiltext{6}{Department of Physics and Astronomy, Arizona State University,
Tempe, AZ 85287}
\altaffiltext{7}{Astronomy Department, California Institute of Technology, 
Pasadena, CA 91125}
\altaffiltext{8}{Institute of Astronomy, University of Cambridge, Cambridge
CB3 0HA, England}
\altaffiltext{9}{Clay Fellow, Harvard-Smithsonian Center for Astrophysics, 60 
Garden St., Cambridge, MA 02138}
\altaffiltext{10}{Department of Astronomy, University of Virginia, 
Charlottesville, VA 22903}
\altaffiltext{11}{Department of Astronomy, University of Wisconsin, Madison, WI
53706}

\begin{abstract}
Ultraviolet and optical images of the face-on spiral galaxies NGC 6753 and NGC 
6782 reveal regions of strong on-going star formation that are associated with 
structures traced by the old stellar populations.  We use these images to 
construct $NUV$---$(NUV - I_{814})$ pixel color-magnitude diagrams (pCMDs) that 
reveal plumes of pixels with strongly varying $NUV$ surface brightness and 
nearly constant $I_{814}$ surface brightness.  The plumes correspond to sharply 
bounded radial ranges, with $(NUV - I_{814})$ at a given $NUV$ surface 
brightness being bluer at larger radii.  The plumes are parallel to both the 
reddening vector and simple model mixtures of young and old populations, thus 
neither reddening nor the fraction of the young population can produce the 
observed separation between the plumes.  The images, and radial 
surface-brightness and color plots indicate that the separate plumes are caused 
by sharp declines in the surface densities of the old populations at radii
corresponding to disk resonances.  The maximum surface brightness of the $NUV$
light remains essentially constant with radius, while the maximum $I_{814}$ 
surface brightness declines sharply with radius.  An $MUV$ image of NGC 6782 
shows emission from the nuclear ring.  The distribution of points in an $(MUV - 
NUV)$---$(NUV - I_{814})$ pixel color-color diagram is broadly consistent with 
the simple mixture model, but shows a residual trend that the bluest pixels in 
$(MUV - NUV)$ are the reddest pixels in $(NUV - I_{814})$.  This may be due to 
a combination of red continuum from late-type supergiants and 
[{\protect\ion{S}{3}}] emission lines associated with {\protect\ion{H}{2}} 
regions in active star-forming regions.  We have shown that pixel mapping is 
a powerful tool for studying the distribution and strength of on-going star 
formation in galaxies.  Deep, multi-color imaging can extend this to studies of 
extinction, and the ages and metallicities of composite stellar populations in 
nearby galaxies.
\end{abstract}

\keywords{galaxies: individual (NGC 6753, NGC 6782) --- galaxies: photometry 
--- galaxies: spiral --- galaxies: structure --- ultraviolet: galaxies}

\section{Introduction}

It has been known since the 1950s that spiral galaxy morphology can be a strong
function of wavelength (\markcite{fzw}Zwicky 1955).  Different bandpasses 
transmit differing amounts of light from young versus old stellar populations, 
and are affected by dust extinction in different ways.  Classical galaxy 
morphology is based on blue-sensitive photographic data, and is thus strongly 
influenced by the distribution of recent star formation and dust.  Over the 
last two decades it has become increasingly clear that the underlying 
distribution of the old stellar mass, as traced by red, or near-infrared (NIR) 
imaging, can be quite different than the $B$-band morphology (e.g., 
\markcite{hs}Hackwell \& Schweizer 1983; \markcite{blo}Block et al.~1994; 
\markcite{fqp}Frogel, Quillen \& Pogge 1996; \markcite{bap}Eskridge et 
al.~2002).  The ability to image galaxies in the vacuum ultraviolet increases 
the lever-arm over which we can study the wavelength dependence of galaxy 
morphology, and can allow us to draw a sharp distinction between UV-bright 
regions and the NIR luminosity distribution.  In the $B$-band, even actively 
star-forming galaxies have a significant contribution to their flux from stars 
as old as a few Gyr (e.g., \markcite{sns}Sage \& Solomon 1989).  By contrast, 
in the vacuum ultraviolet nearby late-type galaxies are dominated by regions of 
current unobscured active star formation.  The spiral arms are prominent, as 
are star-forming rings (if present), and the bulges are weak or not visible at 
all (e.g., \markcite{ku0}Kuchinski et al.~2000, \markcite{ku1}2001; 
\markcite{mea}Marcum et al.~2001; \markcite{pzp}Windhorst et al.~2002).  The 
distribution of current, unobscured, active star formation in a galaxy, 
provided by the ultraviolet morphology, offers us a map of the dynamically 
interesting places in the disk.  The distribution of the old stars, from red or 
NIR imaging, tells us about the underlying gravitational potential.

One means of examining multicolor imaging data of unresolved sources is to
examine the properties of the image stack on a pixel-by-pixel basis.  This
approach has recently been advocated by \markcite{aea}Abraham et al.~(1999), 
but was first discussed by \markcite{b86}Bothun (1986).  \markcite{aea}Abraham 
et al.~(1999) made pixel color-color diagrams, and found relationships between 
location in these color-color diagrams and physical location within the target 
galaxy.  They used {\it Hubble Space Telescope} (HST) images of faint 
intermediate-redshift galaxies in the Hubble Deep Field, taken with the 
Wide-Field and Planetary Camera 2 (WFPC2).  In this paper, we bring the 
technique back to $z=0$ and apply it to the study of star formation in nearby 
spirals.  

Our data include moderately deep HST/WFPC2 exposures for the spiral galaxies 
NGC 6753 and NGC 6782 in two bandpasses (F300W or $NUV$, and F814W or 
$I_{814}$).  We also have a shallow F255W (or $MUV$) image of NGC 6782.  As we 
have moderately deep images in only two bandpasses, we adopt the approach of 
\markcite{b86}Bothun (1986), who first made a pixel color-magnitude diagram 
(pCMD) for NGC 4449.  The advantages we have over Bothun's earlier work derive 
from the special attributes of the HST/WFPC2 combination.  As HST is in orbit, 
we can take data in the vacuum ultraviolet as well as the red part of the 
optical bandpass.  This gives us a larger spectral baseline to work with.  The 
other, somewhat paradoxical, advantage is that WFC images are undersampled.  
This has the consequence that individual pixels are statistically independent.  
Thus well-aligned multiband images with WFPC2 provide a set of statistically 
independent pixels, each of which contains information on the stars within it.  
Each pixel contains a mix of different stellar populations, dust and gas, and 
can be treated as a distinct physical unit, similar in some ways to the study 
of star clusters.  The number and band-passes of the available images 
determines the sorts of information a given data-set will probe.  The 
combination of $NUV$ and $I_{814}$ images we have available is primarily 
sensitive to the relative fractions of young and old stellar populations.

Traditional studies of the structure and contents of galactic disks are based 
on radial, azimuthally averaged, color and surface-brightness profiles (e.g.,
\markcite{k77}Kormendy 1977; \markcite{k85}Kent 1985; \markcite{wnk}Walterbos 
\& Kennicutt 1987; \markcite{deJ}de Jong 1996; \markcite{bba}Baggett, Baggett 
\& Anderson 1998; \markcite{deG}de Grijs 1998; \markcite{pea}Peng et al.~2002). 
This washes out any small-scale effects (due, for instance, to spiral arms and 
interarm regions).  We combine these more traditional approaches with the 
pixel-mapping technique to gain a fuller picture of the distribution of stellar 
populations in the disks of spirals.

Our targets for this study are both bright, nearly face-on early- to mid-type 
spiral galaxies.  Because of their small angular size (they fit well within the 
WFPC2 field of view), they are the only two galaxies from the Ohio State 
University Bright Spiral Galaxy Survey sample (\markcite{bap}Eskridge et 
al.~2002) to have been included in the HST UV imaging sample of 
\markcite{pzp}Windhorst et al.~(2002).  Both galaxies were previously known to 
be UV sources (\markcite{fst}Bowyer et al.~1995).  The basic properties of the 
target galaxies are given in Table 1.  In \S 2 we describe the observations in 
detail, discuss steps we have taken to address the red leak of the UV filters, 
and present our adopted photometric calibration.  We present the $NUV$---$(NUV 
- I_{814})$ pCMDs for both galaxies in \S 3.  In \S 4 we show the pixel 
color-color diagram including the $MUV$ data for NGC 6782.  We discuss these 
results in \S 5.  In \S 6 we present our conclusions, and suggestions for 
further research.

\section{Observations and Data Reduction}

The data for this study are a set of UV and optical images obtained with the
WFPC2 on board HST.  For both NGC 6753 and NGC 6782, we have HST WFPC2 images 
taken through the F300W and F814W filters.  For NGC 6782 we also have an image 
obtained through the F255W filter.  Details of the observations are given in 
Table 2.  The nuclei of the target galaxies were placed on the WF3 chip.  In 
Figure 1 we show $\approx 50'' \times 50''$ sections of the WF3 images, 
centered on the galaxy nuclei.  These data were obtained as part of the HST 
program 8645, ``A Survey of Mid-UV Morphology of Nearby Galaxies: Galaxy 
Structure and Faint Galaxy Evolution'' (R.~Windhorst PI).  Details of the 
program and data reduction are given in \markcite{pzp}Windhorst et al.~(2002).

\subsection{Red Leak Correction}

The F300W filter has significant secondary throughput from $\sim$6600\AA\ to 
$\sim$1$\mu$m.  Thus, if one wishes to do good quantitative work with WFPC2 
F300W images, one must consider and correct for the effects of red leak.  One 
of our technical justifications for obtaining images with the F814W filter was 
to allow for good red-leak correction to the $NUV$ images; the band-pass of the 
F814W filter is very similar to that of the F300W red leak (see 
\markcite{pzp}Windhorst et al.~2002).  

As both NGC 6753 and NGC 6782 are early-type spirals, they are likely to be 
brighter in the wavelength regime of the red leak than they are in the NUV.  
One can make a very simple worst-case estimate of the red leak as follows:  The 
red leak cannot account for any more than all of the counts in any given area 
of a few pixels of an $NUV$ image.  One can therefore take the $I_{814}$ image, 
and iteratively scale it until one finds a maximum possible red-leak 
contamination on these grounds (that is when all of the counts in any region of
the $NUV$ image are actually due to red leak).  We note that this is almost 
certainly an overestimate of the red leak, as the old stars that dominate the 
emission in the $I_{814}$ image will also emit some NUV photons.  But even in 
this most extreme case, there is no substantial qualitative difference between 
the uncorrected and corrected $NUV$ frames for our targets.  To understand why
this is so, note in Figure 1 that the light profiles of both galaxies are 
strongly peaked in the $I_{814}$-band, but only weakly peaked in the $NUV$.  
Thus, when the $I_{814}$-band images are scaled with the constraint that the
subtracted image has no significant negative regions, this has the effect that
the subtracted images reach zero counts in the dust lanes around the nuclei,
while the nuclei themselves, and the surrounding regions of star formation are
nearly unaffected.

One can attempt a more accurate assessment of the red leak in a number of ways.
One relatively simple method is to fold model galaxy spectra through the
response functions of the HST+WFPC2+F814W and F300W combinations, and measure
the expected red leak for various input spectra.  We undertook this experiment
using spectral models from the \markcite{b93}Bruzual \& Charlot (1993) atlas, 
as implemented in the {\sl synphot} package (\markcite{bns}Bushouse \& Simon
1998) within STSDAS\footnotemark\footnotetext{STSDAS was developed at the Space 
Telescope Science Institute (STScI is operated by the Association of
Universities for Research in Astronomy, Inc., for NASA).}.  We proceeded as 
follows:  For our two targets, we measured the total counts in DN (all our 
data were obtained at a gain of 7 electrons/DN) in the F814W frames within a 
circular aperture of a radius of 20 pixels centered on the nucleus.  The 
smooth, old stellar populations dominate both galaxies in this region.  We then 
used a set of \markcite{b93}Bruzual \& Charlot (1993) models to derive 
$I_{814}$-band magnitudes consistent with the measured F814W count rates.  The 
models we used were a pure single burst model with an age of $10^{10}$ yr, a 
$10^{10}$ year old model with an exponentially decreasing star-formation rate 
and an e-folding time of 1 Gyr, and a $10^{10}$ year old model with a constant 
star-formation rate.  For each of these models, we then used our derived 
$I_{814}$-band magnitudes to predict the F300W count rate in the wavelength 
range 6000\AA\ to 12000\AA, and used the integration times of the $I_{814}$ and 
$NUV$ images to derive the appropriate scale factors to apply to the $I_{814}$ 
images.  We then subtracted the scaled $I_{814}$ image from the $NUV$ image.  
Details of these models are given in Table 3, where row 1 shows the observed 
$I_{814}$ count rate (in DN) in the 20-pixel aperture, row 2 shows the 
observed $NUV$ count rate in the same aperture, row 3 shows the implied 
$I_{814}$-band aperture magnitude, row 4 shows the red leak $NUV$ count rate 
for the model, and row 5 shows the fraction of the observed counts in the 
aperture that are due to red leak.

We have also made empirical red-leak corrections, using template spectra from
the Kinney atlas (\markcite{kea}Kinney et al.~1996).  We use their Sa (for NGC
6782) and Sb (for NGC 6753) template spectra.  Our procedure for the Kinney 
atlas templates is the same as outlined above for the model templates, and the 
results are included in Table 3.  Both the \markcite{b93}Bruzual \& Charlot 
(1993) model spectra and the \markcite{kea}Kinney et al.~(1996) empirical 
templates give consistent results.  The F300W red leak does not have any 
qualitative effect on the NUV morphology of our target galaxies.  Even in the 
nuclei, the red leak accounts for $<$10\% of the counts in the NUV images of 
either galaxy.  Further, the range in red-leak correction factors for all the 
templates considered is only a fraction of a percent in both galaxies.  We 
conclude that the F300W red leak does not have any qualitative effect on the 
NUV morphology.

For the F255W image of NGC 6782, we make no attempt at red-leak correction.  
First, the F255W redleak is at $\sim$4000\AA, and we have no HST images at this
wavelength.  Second, the total counts in a 20 pixel radius aperture, centered 
on the nucleus are consistent with 0.  Thus there simply is no room in the data
for a significant red leak.  We note that the WFPC2 Handbook 
(\markcite{wf2}Biretta et al.~2000) also shows that the F255W red leak is also 
significantly smaller than the F300W red leak.

\subsection{Photometric Calibration}

We adopt the photometric calibration from the STScI documents archive 
({\tt http://www.stsci.edu/documents/dhb/web/c32\_wfpc2dataanal.fm1.html}),
dated 01 July 1998.  The adopted WF3 gain-7 zeropoints are given in Table 4.  
To compute magnitudes from a given count rate, we adopt the relation
$$m = ZP -2.5 \times \log(DN / t_{exp}) \eqno(1)$$
where $DN$ is the total counts, $t_{exp}$ is the exposure time, and $ZP$ is the
zeropoint for the filter.  These calibrations are based on those given by
\markcite{zps}Holtzman et al.~(1995), with corrections due to the aging of the
detectors.  Our photometry is thus on the Vegamag system.

The basic reduction of the data in this paper is described in detail in 
\markcite{pzp}Windhorst et al.~(2002).  An issue that requires further 
attention for this study is sky subtraction.  We measured the sky levels in the 
original images after the STSDAS ``On-The-Fly'' pipeline calibration, but 
before cosmic-ray clipping (the cosmic-ray clipping routine must add small 
constants to each image to bring all images onto one common sky).  Table 4
gives our measured sky level in each band in units of magnitude per square
arcsecond. 

\section{UV--Optical Pixel CMD}

We use a pixel-mapping technique (\markcite{b86}Bothun 1986; 
\markcite{aea}Abraham et al.~1999) to examine the data.  As we have only two 
moderately deep images per galaxy, we present this information in the form of 
pCMDs.  Figure 2 shows the $NUV$---$(NUV - I_{814})$ pCMDs for NGC 6753 and 
NGC 6782.  There are two or three bright foreground stars in each image that 
can create spurious features in the pCMDs (one is visible in the lower right 
corner of the image sections of NGC 6782 in Fig.~1).  After masking out these 
bright stars, we plot all pixels with at least three counts (DN) in both the 
sky-subtracted, red-leak-corrected $NUV$ and the $I_{814}$ images.  In 
practice, the signal limit is always reached in the $NUV$ image first.  The 
WFPC2 Handbook (\markcite{wf2}Biretta et al.~2000) gives an effective gain of 
$6.90 \pm 0.32$ electrons per DN, and a read-noise of 5.22 electrons for WF3 at 
gain-7.  Our data were obtained at a detector temperature of $-88~^{\circ}C$, 
giving a dark current of 0.0045 electrons per second.  These, plus our measured 
sky brightness in the $NUV$ (see Table 4) and exposure times (see Table 2) 
allow us to compute the signal to noise ratio (SNR) in the $NUV$ for a limiting 
number of 3 DN per pixel.  For NGC 6753 (two 950 second exposures), this works 
out to a $SNR \approx 4.0$.  For NGC 6782 (three 500 second exposures), we have 
$SNR \approx 5.0$.

On the right margins of Figure 2 we show the absolute magnitude scale for the
$NUV$ data for our assumed distances (see Table 1).  For these distances, a WF3 
pixel subtends approximately 20 pc for NGC 6753 and 25 pc for NGC 6782.  The 
data have been corrected for foreground extinction using the extinction map of 
\markcite{sfd}Schlegel, Finkbeiner \& Davis (1998), and the reddening law of 
\markcite{rip}Cardelli, Clayton \& Mathis (1989).  The arrows in Figure 2 show
reddening vectors for one magnitude of absorption in $m_{NUV}$, using the 
\markcite{rip}Cardelli et al.~(1989) reddening law.  Interior to the left-hand
margins of Fig.~2 we show the apparent surface brightness in magnitudes per
square arcsecond.

There are a number of clear features in the pCMDs.  Plumes of very red 
pixels (shown as black points in Fig.~2) extend up to $M_{NUV} \approx -10$ in 
both galaxies.  For NGC 6753, this plume is at $(NUV - I_{814}) \approx 3.5 \pm 
0.25$, whilst for NGC 6782 it is at $(NUV - I_{814}) \approx 3.3 \pm 0.3$.  
Although this difference in color is not statistically significant, we 
speculate that it may be due to a slightly higher internal extinction in NGC 
6753, as NGC 6753 is a later-type galaxy than NGC 6782.  These red plumes are 
likely to be due to areas that are dominated by the old stellar populations, 
with little or no on-going massive star formation (see \S 5.3 below).  

Both systems show two separate and parallel blue plumes in the pCMDs, 
separated by roughly a magnitude in $(NUV - I_{814})$ color.  These plumes 
extend up to $M_{NUV} \approx -12$ in both galaxies, although they cease to be 
distinct from one another brighter than $M_{NUV} \approx -10.5$ for NGC 6753.  
In both galaxies, we shall refer to the redder of the blue plumes as ``blue
plume 1'' (shown as red points in Fig.~2), and the bluer of the blue plumes as 
``blue plume 2'' (shown as green points in Fig.~2).  In Fig.~2a, the horizontal 
line at $NUV=22.2$ marks the bright limit at which the blue plumes appear
seperate.  Points above this line are plotted in teal in Fig.~2a.  The relative 
density of points in the two plumes is very different in the two galaxies.  In 
NGC 6753, blue plume 1 is the more densely populated, whilst in NGC 6782 it is 
blue plume 2.  The plumes run nearly parallel to the extinction vector.  This 
means that the distribution of pixels within a given plume may be driven 
entirely by the amount of extinction in the direction of that pixel.  It also 
means that the separation between the plumes cannot be an extinction effect.

There are also collections of pixels blueward of the prominent blue plumes in 
both galaxies (show as blue points in Fig.~2).  In NGC 6782, this population is 
clearly separated in $(NUV - I_{814})$ from blue plume 2.  In NGC 6753, it 
forms a blue extension of blue plume 2, and is not clearly separated in color.  
Finally, in NGC 6753 there is a spur emerging from the blue edge of the red 
plume (shown as magenta points in Fig.~2a), at $(NUV - I_{814}) \approx 
2.5$---3.0, extending up to $M_{NUV} \approx -9.5$, and running parallel to the 
blue plumes.

We have added solid lines to Fig.~2 to demarcate these regions of the pCMDs.  
We have also used these boundaries to separate the pixels according to
physical location in their host galaxies.  We show these distributions in 
Figure 3 for NGC 6753 and Figure 4 for NGC 6782, with the same color coding as
in Fig.~2 (the red-plume pixels plotted in black, the blue-plume-1 pixels 
plotted in red, the blue-plume-2 pixels plotted in green, and the very blue 
pixels plotted in blue).  A truly remarkable pattern emerges:  The features 
that are isolated in the pCMDs represent very sharply bounded radial regimes in 
both galaxies, and in both cases there is a strict progression in color, with 
the reddest features in the pCMDs coming from the centers of the galaxies, and 
the bluest features from the largest radii.  In NGC 6753, blue plume 1 
corresponds to pixels in the inner ring (\markcite{csr}Buta 1995).  The 
blue-plume-2 pixels are from the spiral arms outside this inner ring.  In NGC 
6782 both the blue plumes correspond to pixels in the nuclear ring 
(\markcite{csr}Buta 1995), the the seperation between the plumes being the 
strong absorption lanes visible in Fig.~1.  The very blue pixels in NGC 6782 
are in the inner ring (\markcite{csr}Buta 1995).  The separation of the blue 
plumes in color space is a very curious feature.  Metallicity gradients, dust 
distribution patterns, and smooth radial declines in $I_{814}$-band surface 
brightness can all produce a bluing trend with increasing radius, but the gaps 
between features in color space indicate that there is more to the story than 
this.  

In Figure 5, we show $NUV$---$I_{814}$ magnitude-magnitude plots (expressed as
magnitudes per pixel) for the two galaxies.  These plots indicate that the 
pixels in a given blue plume have a nearly constant $I_{814}$-band surface 
brightness, and a large range in $NUV$ surface brightness.  Conversely, in both 
galaxies the two blue plumes have nearly the same range in $NUV$ surface 
brightness, and are seperated by about a magnitude in $I_{814}$-band surface 
brightness.  Thus the separation of the blue plumes is due to drops in the 
$I_{814}$-band surface brightness at particular radii.  Further, as the $NUV$ 
surface brightness distribution is the same in the two plumes, this means the 
drops in the $I_{814}$-band surface brightness have essentially no effect on 
the absolute strength of the on-going unobscured star formation.  We speculate 
that the abrupt changes in the $I_{814}$-band surface brightness are due to 
dynamical effects in the disks of the galaxies that govern the long-term global 
star formation in the disks.  We discuss this matter further in \S 5, below.  

\section{MUV -- NUV -- I Color Map for NGC 6782}

As we obtained an $MUV$ image of NGC 6782, in addition to the $NUV$ and 
$I_{814}$-band images, we examined the $(MUV - NUV)$---$(NUV - I_{814})$ 
color-color plot.  Unfortunately, there are very few pixels with a significant 
number of counts in the $MUV$ image.  All the photons are from the star-forming 
ring (see Fig.~1).  We restrict our analysis to only those pixels with at least 
3 DN in the the $NUV$ ($SNR \ga 5.0$) and $I_{814}$-band images, and 2 DN ($SNR 
\ga 3.6$) in the $MUV$ image (the formal limit of 3 DN on the $I_{814}$ image 
is never reached in practice).  When we look at the distribution of these 
pixels in physical space, we find that they are from three spatially distinct 
regions in the star-forming ring.  One is in the region of the ring that 
corresponds to blue plume 1 (the patch on the inner edge of the visible ring, 
at about 2 o'clock, in the $MUV$ image shown in Fig.~1), while the other two 
are in the region corresponding to blue plume 2 (the patches at about 5 and 10 
o'clock in the $MUV$ image).  The color-color plot, shown in Figure 6, reveals 
that the ring has a foreground-corrected $(MUV - NUV) \approx -0.25 \pm 0.5$.  
The different symbols indicate which patch the pixels come from, with the 
red crosses corresponding to the pixels from the inner part of the ring (the 
patch at about  2 o'clock, in the $MUV$ image), and the green triangles and 
circles corresponding to the pixels from the outer part of the ring 
(respectively, the patches at about 5 and 10 o'clock in the $MUV$ image).  The 
crosses are systematically redder in $(NUV-I_{814})$ than the circles and 
triangles, as would be expected given that they come from the redder plume.  
The arrow in the upper right corner of Fig.~6 is a reddening vector for one 
magnitude of absorption in $m_{NUV}$, using the \markcite{rip}Cardelli et 
al.~(1989) reddening law.  The thin line shows the model color vector of a 
mixture of young and old populations discussed in \S 5.3, below.

There is evidence for a trend with $(NUV - I_{814})$ color such that the bluest 
pixels in $(NUV - I_{814})$ are the reddest pixels in $(MUV - NUV)$ (that is,
in the opposite sense to the model color vector).  \markcite{m88}Majewski 
(1988) saw a similar trend in his optical and NIR data for the starburst ring 
galaxy 52W036.  He noted that the bluest pixels in $(U-B)$ were the reddest 
pixels in $(V-K)$.  For our data, the Spearman and Kendall rank correlation 
tests both indicate that this trend is significant at greater than the 99.9\% 
level for the full sample (83 data points).  If we restrict the test to just 
the pixels from blue plume 2 (56 points), the result holds with even higher 
significance.  The trend breaks down when we consider only the pixels from blue 
plume 1 (27 points).  This may be due to the small number of pixels detected in 
blue plume 1, or it may be that the trend is an astrophysical one that is 
driven by the young stellar population.  In this case, the trend could 
disappear for blue plume 1 due to increased dilution from the old stellar 
population.

While we do not feel our data warrant a detailed investigation of this point,
two possible causes for the observed trend are contamination of the $I_{814}$
fluxes by the emission lines of [{\protect\ion{S}{3}}] from 
{\protect\ion{H}{2}} regions (e.g., \markcite{g89}Garnett 1989; 
\markcite{cdt}Castellanos, D\'{\i}az \& Terlevich 2002) and the presence of a 
significant number of late-type supergiants in the most actively star-forming 
regions (e.g., \markcite{sch}Schaller et al.~1992; \markcite{r98}Rhoads 1998).  
Both of these will lead to enhanced red emission in the regions of the youngest 
stellar populations.  \markcite{m88}Majewski (1988) suggested that his result 
was due to the presence of late-type supergiants as well.  This interpretation 
is bolstered by a consideration of the integrated colors of star clusters in 
the Magellanic Clouds from the study of \markcite{pfc}Persson et al.~(1983).  A
plot of $(U-B)$ against $(V-K)$ shows a U-shaped distribution of points, with
the reddest clusters in $(V-K)$ having either the reddest $(U-B)$ colors (for
the oldest clusters) or the bluest $(U-B)$ colors (for the youngest clusters).
The $K$-band also contains the Br$\gamma$ (2.16$\mu$m) emission line, which 
could also lead to (or enhance) the observed trends.

\section{Discussion}
\subsection{Pixel CMDs}

The technique of using pCMDs to study the structure and stellar 
populations of galaxies is not common, but we are not the first to use it.  
\markcite{b86}Bothun (1986) presented an analysis of a $B$---$(B-R)$ pCMD
of NGC 4449.  His data show a similar blue plume structure to that which we 
find for NGC 6753 and NGC 6782.  There is little evidence for a red plume in 
Bothun's data, but NGC 4449 is a much later-type galaxy than our targets, and 
thus has a relatively more dominant young population.  Also, Bothun's data are 
in the $B$- and $R$-bands, which do not provide as clean a spectral 
discriminant between hot and cool stars as do our $NUV$- and $I_{814}$-band 
data.  \markcite{b86}Bothun (1986) found a spatial segregation in color, 
analogous to what we see in our data (see his Fig.~8), but in the opposite 
sense:  the bluest pixels in NGC 4449 are along the central star-forming ridge
(see the image in \markcite{cag}Sandage \& Bedke 1994), with increasingly red 
pixels at larger radii.  This difference again follows from the much later 
Hubble type of NGC 4449 compared to our targets:  NGC 4449 is an irregular, 
with a central band of vigorous star formation.  Our earlier-type spirals both 
have central bulges, with no significant current central star formation.

As noted in \S4, above, \markcite{m88}Majewski (1988) presented a 
$(U-B)$---$(V-K)$ pixel color-color map of the starbursting ring galaxy 
associated with the radio source 52W036, and noted that the reddest pixels in
$(V-K)$ were the bluest pixels in $(U-B)$.  He also noted the general trend of
increasingly blue color (for the $(U-B)$ color, in his case) with increasing 
radius.  He ascribed this to a radial age gradient in the starbursting ring.

The use of pCMDs was also discussed by \markcite{kaw}Kron, Annis \& Wilhite 
(2000).  They present $g$---$(g-r)$ pCMDs for a spiral (NGC 4030) and an 
elliptical (NGC 4753).  Despite the smaller color baseline and poorer spatial 
resolution of their data, they find structures similar to what we see in our 
data.  They also note that pixels from a given blue plume in NGC 4030 all come 
from the same annulus.  Thus they see the same spatial correlation between 
location and color that we do, and reach the same basic conclusions about its 
origin.  

\subsubsection{Comparison with Radial Profiles}

One measure of the importance of ongoing star formation in a disk is an
estimate of how much of the disk is currently populated by young stars.  The 
fraction of pixels in the inner ring of NGC 6753 (the red zone in Fig.~3) that 
are above a minimum threshold of 3 DN in the $NUV$ is about 60\%.  For the 
nuclear ring of NGC 6782 (the red and green zones in Fig.~4) the fraction of 
pixels above this threshold is about 50\%.  If we extrapolate the relationship 
between $NUV$ and $I_{814}$ surface brightness for the red plumes from Fig.~5, 
we can estimate that the underlying old populations in these regions will have 
$NUV \approx 26$.  This works out to less than 1 DN in the $NUV$.  Thus a 
threshold of 3 DN selects pixels in which the $NUV$ is dominated by the 
unobscured young population.  This implies that at least 50-60\% of the disk 
surface area is engaged in on-going star formation in the most actively 
star-forming radial zones in our targets.  This is really a lower limit, as it 
ignores the (obviously present) effects of extinction. 

A more traditional way of examining the properties of galaxy disks is the use
of radial surface-brightness and color profiles.  In Figure 7 we show
$NUV$ (circles) and $I_{814}$-band (crosses) surface-brightness profiles, and 
$(NUV - I_{814})$ color profiles for NGC 6753.  The profiles were generated 
using elliptical annuli, with the (assumed constant) ellipticities measured 
from the F814W image at a radius of about 50$''$ from the nucleus.  NGC 6753 
shows a peak in $I_{814}$ surface brightness, and then a general decline with 
radius.  There are changes in slope in the surface-brightness profile at about 
10$''$ and at about 18$''$.  The rate of decline of the $I_{814}$ surface 
brightness increases at just beyond these radii.  The $NUV$ surface brightness 
profile shows strong enhancements at radii just inside the slope changes of the
$I_{814}$ surface brightness profile.  These show up as strong blueward peaks 
in the color profile, and correspond to the two blue plumes in the pCMD.  The 
feature at $\sim$10$''$ corresponds to the red points in Fig.~3, and the 
feature at $\sim$18$''$ is the strong spiral arms plotted in green in Fig.~3.

Figure 8 shows the same plots for NGC 6782.  The $I_{814}$ (crosses) 
surface-brightness profile of NGC 6782 shows less structure than that of NGC 
6753.  There is one clear break at about 20$''$.  The $NUV$ (circles)
surface-brightness profile has a much stronger enhancement at this radius, and 
the color profile has a strong blue feature here.  The $NUV$ surface-brightness 
profile also has a strong enhancement at about 5$''$, that shows up as a 
dramatic blue spike in the color profile.  These are due to the two rings of 
star formation, with the feature at $\sim$5$''$ plotted in Fig.~4 as red and 
green points, and the feature at $\sim$20$''$ plotted in Fig.~4 as blue points.

The behavior of the radial profiles of both galaxies supports the conclusion 
that we drew from Fig.~5:  Regions of strong, unobscured, on-going star
formation (as traced by localized enhancements in the $NUV$ surface brightness)
are associated with regions of increased rate of decline in the $I_{814}$-band 
surface brightness.  It is these discontinuities in the $I_{814}$-band surface 
brightness that drive the separations between the plumes we see in the pCMDs.  
The power of a pixel analysis is that it demonstrates that the range of $NUV$ 
surface brightness is the same in each region of strong star formation (see
Fig.~5).  That is, the maximum rate of localized star formation is not a strong 
function of the underlying distribution of old stellar mass.  This point is
lost in a traditional profile analysis.

\subsubsection{Binned Pixel CMDs}

Our pCMDs are noisy.  In particular, they suffer from obvious discreteness
noise due to the low count rate in the NUV data.  Binning the data makes the
trade off of lowering the spatial resolution while allowing us to probe fainter
surface brightness levels at a given signal to noise.  We made 2$\times$2 and 
3$\times$3 binned images of both galaxies, and generated pCMDs from the 
binned images.  We imposed a limit of $SNR \ga 3$ per pixel on the binned 
pCMDs.  For NGC 6753 this translates to a minimum of 4 DN per 2$\times$2 binned 
pixel and 6 DN per 3$\times$3 binned pixel.  This means that the 2$\times$2 
binned image probes roughly 1.2 magnitudes fainter in surface brightness than 
does the unbinned image, while the 3$\times$3 binned image goes roughly 1.6 
magnitudes fainter.  For NGC 6782 the limits are 3 DN per 2$\times$2 binned 
pixel and 5 DN per 3$\times$3 binned pixel.  This means that the 2$\times$2 
binned image probes roughly 1.5 magnitudes fainter in surface brightness than 
does the unbinned image, while the 3$\times$3 binned image goes roughly 1.8 
magnitudes fainter.  We present the binned pCMDs in Figure 9.  As in Fig.~2,
interior to the left-hand axis, we plot the surface brightness in magnitudes
per square arcsecond.  Fig.~9a,b show
the 2$\times$2 and 3$\times$3 binned binned pCMDs for NGC 6753, respectively,
while Fig.~9c,d show the the 2$\times$2 and 3$\times$3 binned binned pCMDs for 
NGC 6782.  NGC 6753 now shows evidence for at least four blue plumes.  The 
improvement in the pCMD for NGC 6782 is less dramatic, but the blue plumes are 
better defined, and extend to much fainter surface brightness than in the 
unbinned data.

The binned pCMDs for NGC 6753 (Fig.~9a,b) show the same basic features at
high surface brightness as does the unbinned pCMD.  The blue plumes are a bit
more diffuse.  At lower surface brightness, some interesting features emerge.
The red plume has a clear faint limit (at $NUV \approx 23.5$ in Fig.~9b), much
brighter than the SNR cut-off imposed.  This limit is the minimum NUV surface
brightness of the old population in the bulge region of NGC 6753.  There is
also a faint limit to the inner blue plume that reflects the minimum NUV 
surface brightness of the inner ring, at $NUV \approx 24$ in Fig.~9b.  Given
our assumed distance (see Table 1), this is roughly the expected apparent 
magnitude of a single O5 main sequence star.  The distribution of pixels 
blueward of the blue plume 2 in Fig.~2a (the blue region in Fig.~4) is a 
coherent plume in Fig.~9a,b.  These pixels are due to the outer star-forming 
regions in the disk.

The binned pCMDs for NGC 6782 (Fig.~9c,d) show the truncation of the red plume,
similar to what is seen in NGC 6753.  They also reveal the benefit of high
angular resolution:  The two distinct blue plumes in Fig.~2b are merged 
together in the binned pCMDs.  However, the star forming knots associated with 
the inner ring (not the nuclear ring!) form a well-developed blue plume in the 
Fig.~9c, running from ($NUV \approx 24.6$, $(NUV - I_{814}) \approx 1.3$) to 
($NUV \approx 22$, $(NUV - I_{814}) \approx -0.4$).  Only the very brightest 
knots associated with this plume show up in the unbinned pCMD.  

\subsection{Disk Dynamics}

The $NUV$ light traces the distribution of on-going unobscured star formation 
in our targets.  In both galaxies, the majority of the on-going star formation 
occurs in small radial ranges.  That is, the current star forming properties of 
the disks are governed by the gravitational dynamics of the disks.  Both NGC 
6753 and NGC 6782 are in the Catalog of Southern Ringed Galaxies 
(\markcite{csr}Buta 1995), in which they are classified as (R$'$)SA(r)b (NGC 
6753) and (R$_1'$)SB(r)0/a (NGC 6782).  Thus there is significant prior work on 
the internal dynamical state of both systems.

In both galaxies, there is no evidence for current star formation in the inner 
parts of the galaxies (the red plumes).  \markcite{bnc}Buta \& Crocker (1993)
have argued that the onset of strong star formation in NGC 6753 (that is, the
inner boundary of the region that is responsible for blue plume 1) occurs at
the inner Lindblad resonance (ILR).  The inner ring in NGC 6753 is the 
structure that marks the spatial boundary between the pixels forming the two 
blue plumes.  \markcite{cea}Crocker, Baugus \& Buta (1996) identify this as the 
location of the 4 to 1 ultraharmonic resonance (UHR).  At this point, the 
long-term star formation rate drops abruptly, and blue plume 2 shows pixels of 
ongoing star formation in the disk outside the UHR.  The very blue pixels show 
the knots of current star formation in the outer ring, which is most likely 
associated with the co-rotation radius (CR).

The bright star-forming ring that is so prominent in our $NUV$ image of NGC 
6782 is not the inner ring, but is actually a third, nuclear ring.  
\markcite{bea}Byrd et al.~(1994) associate the nuclear ring with the ILR.  The 
blue plumes correspond to the inside and outside of the dust lane associated 
with this nuclear ring that surrounds the nuclear bar.  The $I_{814}$-band 
surface brightness drops abruptly at the radius of the dust lane, and outside 
this region current star formation ceases entirely.  The outer blue plume is 
due to a small number of star forming knots in the inner ring that is 
associated with the terminus of the inner bar.  The bar is visible in our 
$I_{814}$ image, and is prominent in ground-based NIR images 
(\markcite{bap}Eskridge et al.~2002).  \markcite{bea}Byrd et al.~(1994) 
associate the inner ring with the UHR.  

\subsection{Stellar Populations}

Our data show evidence for dynamically triggered on-going star formation added
on to the accumulated older stellar populations of the target galaxies.  
In order to put this interpretation on a firmer footing, we compare our 
data to model properties of single stars, and composite stellar 
populations, and to observations of young stellar clusters.

\subsubsection{Comparisons with Stellar Models}

\markcite{rea}Romaniello et al.~(2002) have published model stellar 
color-temperature data for a number of HST/WFPC2 bandpasses.  Figure 10 shows 
the predicted relationship between effective temperature and our 
$(NUV-I_{814})$ color for both main-sequence ($\log g = 4.5$) and giant ($\log 
g = 3$) stars.  The reddest colors observed in our pCMDs for both NGC 6753 and 
NGC 6782 are $(NUV - I_{814}) \approx +4$.  The models of 
\markcite{rea}Romaniello et al.~(2002) indicate that such a color is 
appropriate for stars with $T_{eff} \approx 5000$ K; early KV or late GIII 
stars.  The turn-off of an old stellar population dominates its UV light, and 
an early K turn-off is appropriate for the stellar population that dominates 
the centers of early- to mid-type spirals.  As this is exactly where the 
red-plume pixels are located, this argues strongly that the red plumes are due 
to the old stellar population.

The bluest colors predicted in the \markcite{rea}Romaniello et al.~(2002) 
models are $(NUV - I_{814}) \approx -3$, for O stars ($T_{eff} \approx 5 \times 
10^4$ K).  The bluest pixels in our pCMD for NGC 6753 are also $(NUV - I_{814}) 
\approx -3$, supporting our argument that the light from these pixels is  
dominated by unobscured on-going massive star formation.  The bluest pixels in 
the pCMD for NGC 6782 are only $(NUV - I_{814}) \approx -1$.  If the light from 
these pixels is dominated by current/recent star formation, that implies the 
maximum stellar temperature in NGC 6782 is only $T_{eff} \approx 1.3 \times 
10^4$ K, corresponding to late B stars.  An alternative explanation is that 
even the bluest pixels in NGC 6782 have a substantial contribution to their 
light from the older, redder stellar population.  If we assume that the bluest 
pixels in NGC 6782 have all their NUV light coming from O stars, the absolute 
magnitudes of those pixels ($M_{NUV} \approx -12$) require about 20-25 O5 stars 
per pixel.  The colors of the pixels are $(NUV - I_{814}) \approx -1$, so $M_I 
\approx -11$.  The \markcite{rea}Romaniello et al.~(2002) models indicate that 
these stars would have an integrated $I_{814}$-band absolute magnitude of 
$M_I(young) \approx -9$.  This means that the total contribution of the 
underlying old population would need to be $M_{I}(old)\approx-10.8$.  Assuming 
this light to be produced by K5III stars leads to the requirement of about 5000 
such stars per pixel.  The predicted integrated $NUV$ absolute magnitude of 
5000 K5III stars is only $M_{NUV}(old) \approx -5.3$; a negligible contribution 
to the total $M_{NUV} \approx -12$.  Thus the O stars account for all the $NUV$ 
light, and the K stars dominate the $I_{814}$-band light.

\subsubsection{Comparisons with Simple Stellar Population Models}

We can approach the problem from a different direction using the spectral
evolution models of \markcite{b93}Bruzual \& Charlot (1993).  The question we
wish to probe is how mixing various amounts of a young stellar population with
an underlying old population will effect the predicted $(NUV - I_{814})$ color 
of the resulting composite stellar population.  We adopt the 
\markcite{b93}Bruzual \& Charlot (1993) 10 Gyr model, with a low-mass cut-off 
of 0.1 $M_{\odot}$ as our old model, and the 1 Myr model, with a low-mass 
cut-off of 0.1 $M_{\odot}$ and high-mass cut-off of 125 $M_{\odot}$ as our 
young model.  We combine these models with variable weights, and convolve the 
combinations with the sensitivity curves for our filters, using {\sl synphot}.  
We picked a constant $I_{814}$-band magnitude, and determined the predicted 
$(NUV - I_{814})$ color for a range of weights.  As we constrain the 
$I_{814}$-band magnitude, the weights reflect the input contributions of the 
young and old populations to the desired $I_{814}$-band magnitude.  The results 
of this experiment are given in Table 5, and shown in Figure 11.  Figure 11a 
shows the data for NGC 6753.  The dotted line is the vector resulting from 
combining a range of weights of young and old populations with the constraint 
that $I_{814} = 22.8$ (roughly the magnitude of the base of blue plume 1).  The 
circles show the knot points, with a pure young population at the top, and an 
order of magnitude decrease in the weight of the young population for each 
circle.  The circle at ($(NUV-I_{814})=1.5$, $NUV=24.3$) results from a young 
population that contributes only $10^{-4}$ of the $I_{814}$ light.  Figure 11b 
shows the same models overlayed on the NGC 6782 pCMD, with the constraint 
shifted to $I_{814} = 21.8$.  The knot point at ($(NUV-I_{814})=1.5$, 
$NUV=23.3$) results from a young population that contributes only $10^{-4}$ of 
the $I_{814}$ light.

The model runs nearly parallel to the blue plumes in NGC 6753.  The model and
the data are a bit more skewed for NGC 6782, but in both cases the loci defined 
by these simple models are good representations of the distributions of the 
points in the blue plumes of the pCMDs.  Thus one can produce the observed
distribution of points in the pCMDs by either simple mixtures of young and old
populations, or by pure young populations and reddening (although the existence 
of an old population follows from the $I_{814}$-band morphologies).  Note that 
this also means, at least in the case of NGC 6753, that the separation between
the plumes is not due to variations in the current level of star formation laid 
atop a constant old population.  This will move a given point along a plume, 
but will not move it from one plume to another.

In Fig.~6, we show the locus of the mixture model, as a solid line, in the 
$(MUV - NUV)$---$(NUV - I_{814})$ color-color plane.  The two large squares 
show the predicted colors for young population fractions of $10^{-4}$ and 
$10^{-3}$ (see Table 5).  The points generally scatter around the model, and 
indicate that a small fraction of on-going star formation plus substantial and 
variable reddening can produce the colors we see.  We note that the generally
redder colors of the pixels from the inner part of the nuclear ring can easily
be due to either reddening or to a larger contribution from the old stellar
population.

\subsubsection{Comparisons with Observations of Young Stellar Clusters}

A final test of our basic picture is provided by the data of 
\markcite{hea}Harris et al.~(2001), who provide F300W and F814W observations of
young star clusters in M83.  They estimate the ages of these clusters to be in
the range 2--50 Myr.  In Figure 12, we show the $(NUV - I_{814})$ colors of the 
M83 starclusters plotted against their ages.  The youngest M83 clusters are 
generally the bluest, and have typical integrated colors of $(NUV - I_{814}) 
\approx -3.5$, consistent with the bluest plumes in NGC 6753.  The oldest 
clusters have colors of $(NUV - I_{814}) \approx -1.5$, more similar to the 
bluest plumes we see in NGC 6782.  These tests show that the redder colors of 
the blue plumes in NGC 6782, compared to NGC 6753, can be due to either an 
older starburst population, or a larger fraction of old stars mixed with a 
zero-age starburst population.  We believe the latter interpretation is more 
likely correct, as the active star formation in NGC 6782 occurs in the very 
inner part of the galaxy, and NGC 6782 is an earlier-type system than NGC 6753.

\section{Summary and Conclusions}

We have analysed HST/WFPC2 $NUV$ and $I_{814}$ images of the face-on spiral
galaxies NGC 6753 and NGC 6782.  We use these images to construct $NUV$---$(NUV 
- I_{814})$ pCMDs for these galaxies.  Our data are a set of aligned HST/WFPC2 
images, with the images for a given galaxy all obtained in the same orbit.  
Thus we have sets of well-aligned, undersampled images.  This is a real 
advantage for pixel-mapping, as it results in essentially statistically 
independent pixels.  One can thus treat each pixel as a pseudo-star cluster.  
This work opens the possibility of studying the detailed interplay between the 
star-formation history and dynamics of nearby disk galaxies.  As the 
$I_{814}$-band light is mainly from the old population, it is a good tracer of 
the stellar mass.  The $I$-band mass to light ratio is between 1 and 2 in Solar 
units, and is not a strong function of age for populations older than 1 Gyr or 
so.  So the $(NUV-I_{814})$ color is, in effect, a measure of the current 
unobscured star formation rate per unit stellar mass.  

Our pCMDs reveal clear, separated plumes of pixels.  Each of these plumes have 
strongly varying $NUV$ surface brightness, and nearly constant $I_{814}$ 
surface brightness.  The maximum and average $NUV$ surface brightness is the
same from plume to plume.  It is the decline with increasing radius of the 
maximum $I_{814}$ surface brightness that causes the separation between the 
plumes.  In both galaxies, the plumes correspond to distinct, sharply bounded 
radial ranges, with the $(NUV - I_{814})$ color at a given $NUV$ surface 
brightness being bluer at larger radii.  These plumes are parallel to both the 
reddening vector, and simple models of mixtures of young and old populations.  
Thus it is unclear from our data if the distribution of points in a given plume 
is primarily due to variations in reddening or in the relative importance of a 
young stellar population.  Conversely, it is clear that neither reddening, nor 
varying the importance of a young population can produce the separations 
between the plumes.  

The $I_{814}$ images, as well as radial surface-brightness and color 
plots indicate that the separate plumes are caused by discontinuities in the
surface density of the old stellar population.  These discontinuities are 
the result of the accumulated history of star formation in the disks of these 
galaxies.  The current star formation, as traced by the $NUV$ images is clearly
driven by dynamical triggers in both galaxies.  NGC 6782 is a particularly
clear example of this since all the current star formation, as traced by the 
$NUV$ light, is associated with the nuclear and inner rings.

We have an $MUV$ image of NGC 6782 that shows emission from the nuclear ring.
We have used this to generate an $(MUV - NUV)$---$(NUV - I_{814})$ pixel 
color-color diagram.  The colors are generally consistent with a model that
mixes a small fraction of a young stellar population with a dominant old
population.  Either variations in the relative strength of on-going star 
formation or reddening can produce the scatter we see in the color-color plot.  
A curious feature of this diagram is that the bluest pixels in $(MUV - NUV)$ 
tend to be the reddest pixels in $(NUV - I_{814})$.  This trend is 
statistically significant based on both the Spearman and Kendall rank tests.  
We speculate that it may be due to a combination of enhanced 
[{\protect\ion{S}{3}}] emission from {\protect\ion{H}{2}} regions and the red 
flux from late-type supergiants associated with very young stellar populations.

The sort of study we have done in this paper is now possible for a substantial
number of galaxies as the result of our Cycle 9 (\markcite{pzo}Windhorst et 
al.~2002) and 10 HST UV imaging programs\footnotemark\footnotetext{Cycle 10 HST 
program 9124, ``Mid-UV Snapshot Survey of Nearby Irregulars: Galaxy Structure 
and Evolution Benchmark'', R.~Windhorst PI}.  A more densely sampled spectral 
energy distribution would allow for more detailed stellar population modelling 
than the admittedly crude attempt we have made here (e.g., \markcite{f85}Frogel
1985).  We are encouraged that the Hubble Heritage Team followed up our imaging 
of NGC 6782 with a multi-band Heritage imaging program.  The addition of just 
the remaining optical bandpasses ($UBVR$) will allow a much more accurate 
disentangling of the various effects of age, extinction, metallicity, and the 
relative importance of different populations than has been possible here.  One 
clear task for the future is to use the Heritage data to investigate the 
stellar populations and effects of dust in the disk of NGC 6782.  

A more ambitious program for the future would be to observe these galaxies with
the NICMOS or WFC3 on HST in order to obtain NIR colors and 2.29 $\mu$m CO
index images.  Broad-band NIR imaging would provide data on the distribution of
the old stellar populations that are essentially free from the effects of dust.
The NIR CO index is an excellent diagnostic for the relative importance of
late-type supergiants in an unresolved stellar population, due to its strong
dependence on surface gravity (\markcite{bfp}Baldwin, Frogel \& Persson 1973). 
Observations of nearby galaxies (\markcite{r98}Rhoads 1998) have shown that
CO index measurements are a potentially powerful tool for studying the 
distribution of recent star formation in galaxies, as they are sensitive to
the distribution of massive supergiants, and are relatively unaffected by dust. 

\acknowledgments
PBE would like to thank the members of the Astronomy Department at Ohio State
University, where this project was begun.  JAF acknowledges the support of NASA 
Headquarters for publication funds while he is on leave there from OSU.  JAF 
thanks Dr.~Sean Solomon for granting him Visiting Investigator status at
DTM/CIW.  LDM acknowledges support from a Clay Fellowship from the
Harvard-Smithsonian Center for Astrophysics.  We thank R.~Pogge for suggesting 
that [{\protect\ion{S}{3}}] emission could contribute to the effect mentioned 
in \S 4.  We also thank R.~Kron for pointing out the S.~Majewski's work on 
pixel-mapping.  This research has made use of NASA's Astrophysics Data System 
Bibliographic Services, and the NASA/IPAC Extragalactic Database (NED) which is 
operated by the Jet Propulsion Laboratory, California Institute of Technology, 
under contract with the National Aeronautics and Space Administration.   This 
work was supported in part by NASA Hubble Space Telescope grants HST-GO-08645* 
and HST-GO-09124* and the NASA Long Term Space Astrophysics grant NAG5-6403 to 
the University of Virginia.

\clearpage

{
\def\tabrule{\noalign{\hrule}}
\def\pz{\phantom{0}}
\def\pb{\phantom{-}}
\def\pd{\phantom{.}}
\baselineskip12pt
\

\centerline{Table 1 -- Basic Properties}
\vskip0.3cm

\newbox\tablebox
\setbox\tablebox = \vbox {

\halign{\pz\pz#\pz\pz&\hfil\pz\pz#\pz\pz\hfil&\hfil\pz\pz#\pz\pz\hfil&\hfil\pz
\pz#\pz\pz\hfil\cr
\tabrule
\noalign{\vskip0.1cm}
\tabrule
\noalign{\vskip0.1cm}

 & NGC 6753 & NGC 6782 & References \cr
\noalign{\vskip0.1cm}
\tabrule
\noalign{\vskip0.2cm}
\ & (R$'$)SA(r)b & (R)SAB(r)a & 1 \cr
$\alpha$(J2000.0) & $\pb 19^h11^m23{^s}\llap.4$ & $\pb 19^h23^m57{^s}\llap.2$ & 
1 \cr
$\delta$(J2000.0) & $-57^{\circ}02'56''$ & $-59^{\circ}55'22''$ & 1 \cr
$m_B$ & $\pb$11.83 & $\pb$11.84 & 1 \cr
$V_{\odot}$ & $\pb$3142 ${\rm km~s^{-1}}$ & $\pb$3736 ${\rm km~s^{-1}}$ & 1 \cr
D & $\pb$42.6 Mpc & $\pb$51.2 Mpc & 1,2 \cr
$M_B$ & $-$21.3 & $-$21.7 & 1,2 \cr
$(B-V)_T$ & 0.83 & 0.92 & 1 \cr
$D_{25}$ & $2\rlap.'5$ & $2\rlap.'2$ & 1 \cr
\noalign{\vskip0.2cm}
\tabrule
}
}
\centerline{ \box\tablebox}

\vskip15pt
1) \markcite{rc3}de Vaucouleurs et al.~1991.  2) Distance derived using the
\markcite{yts}Yahil, Tammann \& Sandage (1977) formalism, and
$H_{\circ}=70~{\rm km~s^{-1}~Mpc^{-1}}$.  

}

\newpage

\baselineskip12pt
\tolerance=500

\def\tabrule{\noalign{\hrule}}
\def\pz{\phantom{0}}
\def\pb{\phantom{-}}
\def\pd{\phantom{.}}
\def\po{\phantom{1}}
\

\centerline{Table 2 - Log of Observations}
\vskip0.2cm

\newbox\tablebox
\setbox\tablebox = \vbox {

\halign{#\pz&\hfil#\pz\hfil&\hfil#\hfil&\hfil\pz#\pz\hfil&\hfil#\hfil\cr
\tabrule
\noalign{\vskip0.1cm}
\tabrule
\noalign{\vskip0.1cm}

 & NGC 6753 & & NGC 6782 \cr
\noalign{\vskip0.1cm}
\tabrule
\noalign{\vskip0.1cm}
 & Date & Exp & Date & Exp \cr
 & ddmmyyyy & sec & ddmmyyyy & sec \cr
\noalign{\vskip0.1cm}
\tabrule
\noalign{\vskip0.2cm}
MUV & & & 22062000 & 3$\times$467 \cr
NUV & 22062000 & 2$\times$950 & 22062000 & 3$\times$500 \cr
$I_{814}$ & 22062000 & 2$\times$160 & 22062000 & 2$\times$130 \cr
\noalign{\vskip0.2cm}
\tabrule
}
}
\centerline{ \box\tablebox}

\newpage

\baselineskip12pt
\tolerance=500

\

\centerline{Table 3a - NGC 6753 Red Leak Corrections}
\vskip0.2cm

\newbox\tablebox
\setbox\tablebox = \vbox {

\halign{#\pz\hfil&\hfil#\pz\hfil&\hfil#\pz\hfil&\hfil#\pz\hfil&\hfil#\pz\hfil\cr
\tabrule
\noalign{\vskip0.1cm}
\tabrule
\noalign{\vskip0.1cm}

 & B\&C old & B\&C burst & B\&C exp & Kinney \cr
\noalign{\vskip0.1cm}
\tabrule
\noalign{\vskip0.2cm}
F814W c-r & 2806.25$\pz$ \cr
F300W c-r & $\pz\pz\pz$9.15$\pz$ \cr
synphot $I_{814}$ & $\pz\pz$13.07$\pz$ & 13.06$\pz$ & 13.06$\pz$ & 13.07$\pz$ 
\cr
F300W red-leak c-r & $\pz\pz\pz$0.68$\pz$ & $\pz$0.68$\pz$ & $\pz$0.70$\pz$ & 
$\pz$0.67$\pz$ \cr
red-leak fraction & $\pz\pz\pz$0.073 & $\pz$0.075 & $\pz$0.077 & $\pz$0.074 \cr
\noalign{\vskip0.2cm}
\tabrule
}
}
\centerline{ \box\tablebox}

\vskip30pt

\centerline{Table 3b - NGC 6782 Red Leak Corrections}
\vskip0.2cm

\newbox\tablebox
\setbox\tablebox = \vbox {

\halign{#\pz\hfil&\hfil#\pz\hfil&\hfil#\pz\hfil&\hfil#\pz\hfil&\hfil#\pz\hfil\cr
\tabrule
\noalign{\vskip0.1cm}
\tabrule
\noalign{\vskip0.1cm}

 & B\&C old & B\&C burst & B\&C exp & Kinney \cr
\noalign{\vskip0.1cm}
\tabrule
\noalign{\vskip0.2cm}
F814W c-r & 2193.85$\pz$ \cr
F300W c-r & $\pz\pz$10.43$\pz$ \cr
synphot $I_{814}$ & $\pz\pz$13.34$\pz$ & 13.33$\pz$ & 13.33$\pz$ & 13.33$\pz$ 
\cr
F300W red-leak c-r & $\pz\pz\pz$0.53$\pz$ & $\pz$0.53$\pz$ & $\pz$0.53$\pz$ & 
$\pz$0.53$\pz$ \cr
red-leak fraction & $\pz\pz\pz$0.051 & $\pz$0.051 & $\pz$0.051 & $\pz$0.051 \cr
\noalign{\vskip0.2cm}
\tabrule
}
}
\centerline{ \box\tablebox}

\newpage

\baselineskip12pt
\tolerance=500

\

\centerline{Table 4 - Adopted WF3 Gain-7 Zero-points and Sky Levels}
\vskip0.2cm

\newbox\tablebox
\setbox\tablebox = \vbox {

\halign{#\pz\hfil&\hfil#\pz\hfil&\hfil#\pz\hfil\cr
\tabrule
\noalign{\vskip0.1cm}
\tabrule
\noalign{\vskip0.1cm}

Filter & Zero-point & Sky Level \cr
\noalign{\vskip0.1cm}
 & Mag. & Mag. / sq.~arcsec \cr 
\noalign{\vskip0.1cm}
\tabrule
\noalign{\vskip0.2cm}
F255W & 17.037 & 21.85 \cr
F300W & 19.433 & 23.19 \cr
F814W & 21.659 & 21.74 \cr
\noalign{\vskip0.2cm}
\tabrule
}
}
\centerline{ \box\tablebox}

\newpage

\baselineskip12pt
\tolerance=500

\

\centerline{Table 5 - Predicted $(NUV - I_{814})$ Colors}
\vskip0.2cm

\newbox\tablebox
\setbox\tablebox = \vbox {

\halign{#\pz\hfil&#\pz\hfil&\hfil#\pz\hfil&\hfil#\pz\hfil\cr
\tabrule
\noalign{\vskip0.1cm}
\tabrule
\noalign{\vskip0.1cm}

Old Weight & Young Weight & $(NUV - I_{814})$ & $(MUV - NUV)$ \cr
\noalign{\vskip0.1cm}
\tabrule
\noalign{\vskip0.2cm}
 1 & 0 & $\pb$3.46 & $\pb$1.42 \cr
 0.99999 & 0.00001 & $\pb$2.99 & $\pb$0.38 \cr
 0.9999 & 0.0001 & $\pb$1.47 & $-$0.29 \cr
 0.999 & 0.001 & $-$0.67 & $-$0.42 \cr
 0.99 & 0.01 & $-$2.10 & $-$0.44 \cr
 0.9 & 0.1 & $-$2.46 & $-$0.44 \cr
 0 & 1 & $-$2.50 & $-$0.44 \cr
\noalign{\vskip0.2cm}
\tabrule
}
}
\centerline{ \box\tablebox}

\clearpage

\figcaption{50$'' \times$50$''$ sections of the WF3 images of NGC 6753 (top) 
and NGC 6782 (middle) in the $NUV$ (left) and $I_{814}$ (right).  The $MUV$ 
image of NGC 6782 is shown on the bottom left.  The arrows in the upper-right
corners of the F814W images show the direction of North, with the line-segments
pointing East.  All images of a given target have the same orientation.  The 
scale bar in the $MUV$ image of NGC 6782 is 10$''$, and applies to all five 
panels.}

\figcaption{$NUV$ --- $(NUV-I_{814})$ pCMDs for a) NGC 6753 and b) NGC 6782. 
The solid lines are discussed in the text.  The right-hand axis labels show the
absolute magnitude per pixel for our assumed distances and foreground 
reddening.  The numbers interior to the left-hand axis show the surface
brightness in magnitudes per square arcsecond.  The colors identify the
same sets of pixels shown in Figs.~3 and 4.  The arrows show the effect of 
$A_{NUV}=1$, following the Cardelli et al.~(1989) reddening law.}

\figcaption{Spatial maps of NGC 6753 showing the locations of the pixels in the
red plume (black points), blue plume 1 (red points), blue plume 2 (green 
points), the very blue smear (blue points), and the $NUV$-brightest pixels
(purple).}

\figcaption{Spatial maps of NGC 6782 showing the locations of the pixels in the
red plume (black points), blue plume 1 (red points), blue plume 2 (green 
points), and the very blue smear (blue points).}

\figcaption{$NUV$ --- $I$ pixel magnitude diagrams for a) NGC 6753 and b) 6782,
using the same color table as Figs.~3 and 4.}

\figcaption{$(MUV-NUV)$--$(NUV-I_{814})$ color-color pixel map for NGC 6782.  
The red crosses show pixels from blue plume 1, while the green circles and 
triangles show pixels from blue plume 2.  The bold arrow shows the effect of 
$A_{NUV}=1$, following the Cardelli et al.~(1989) reddening law.  The solid 
line shows the effect of mixing small amounts of a young population with an 
underlying old population.  The large squares show the knot points given in 
Table 5.}

\figcaption{Top: $I_{814}$ and $NUV$ surface-brightness profiles for NGC 6753.
The crosses and left-hand label give the $I_{814}$ surface brightness, and the
circles and right-hand label give the $NUV$ surface brightness.  Bottom:  
$(NUV-I_{814})$ color profile for NGC 6753.}

\figcaption{Top: $I_{814}$ and $NUV$ surface-brightness profiles for NGC 6782.
The crosses and left-hand label give the $I_{814}$ surface brightness, and the
circles and right-hand label give the $NUV$ surface brightness.  Bottom:  
$(NUV-I_{814})$ color profile for NGC 6782.}

\figcaption{Binned $NUV$ --- $(NUV-I_{814})$ pCMDs for a) NGC 6753 with 
2$\times$2 pixel binning, b) NGC 6753 with 3$\times$3 pixel binning, c) NGC 
6782 with 2$\times$2 pixel binning, d) NGC 6782 with 3$\times$3 pixel binning.
In all cases, the numbers interior to the left-hand axis show the $NUV$ surface
brightness in magnitudes per square arcsecond.}

\figcaption{Model ($NUV-I_{814})$ colors for stars, plotted against stellar 
effective temperature.  The solid line shows the predicted relationship for 
main sequence stars, and the dotted line for giants.}

\figcaption{$NUV$ --- $(NUV-I_{814})$ pCMDs for a) NGC 6753 and b) NGC 6782, as 
in Figure 2.  The dotted lines show the effect of mixing small amounts of a 
young population with an underlying old population.  The circles show the knot 
points given in Table 5.}

\figcaption{Dereddened $(NUV-I_{814})$ colors plotted against log(age) for star 
clusters in M83, from Harris et al.~(2001)}

\newpage
\plotone{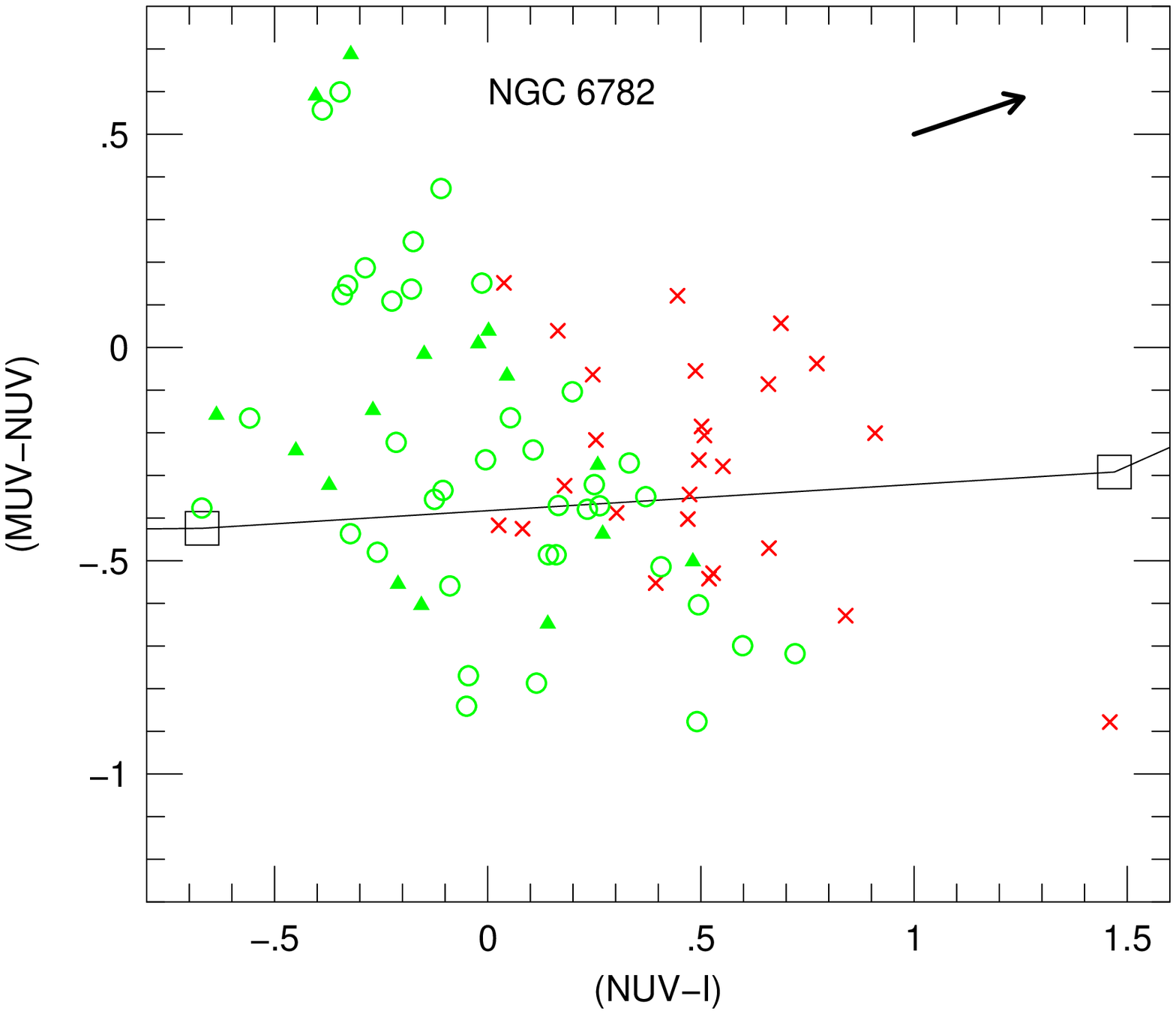}
\newpage
\plotone{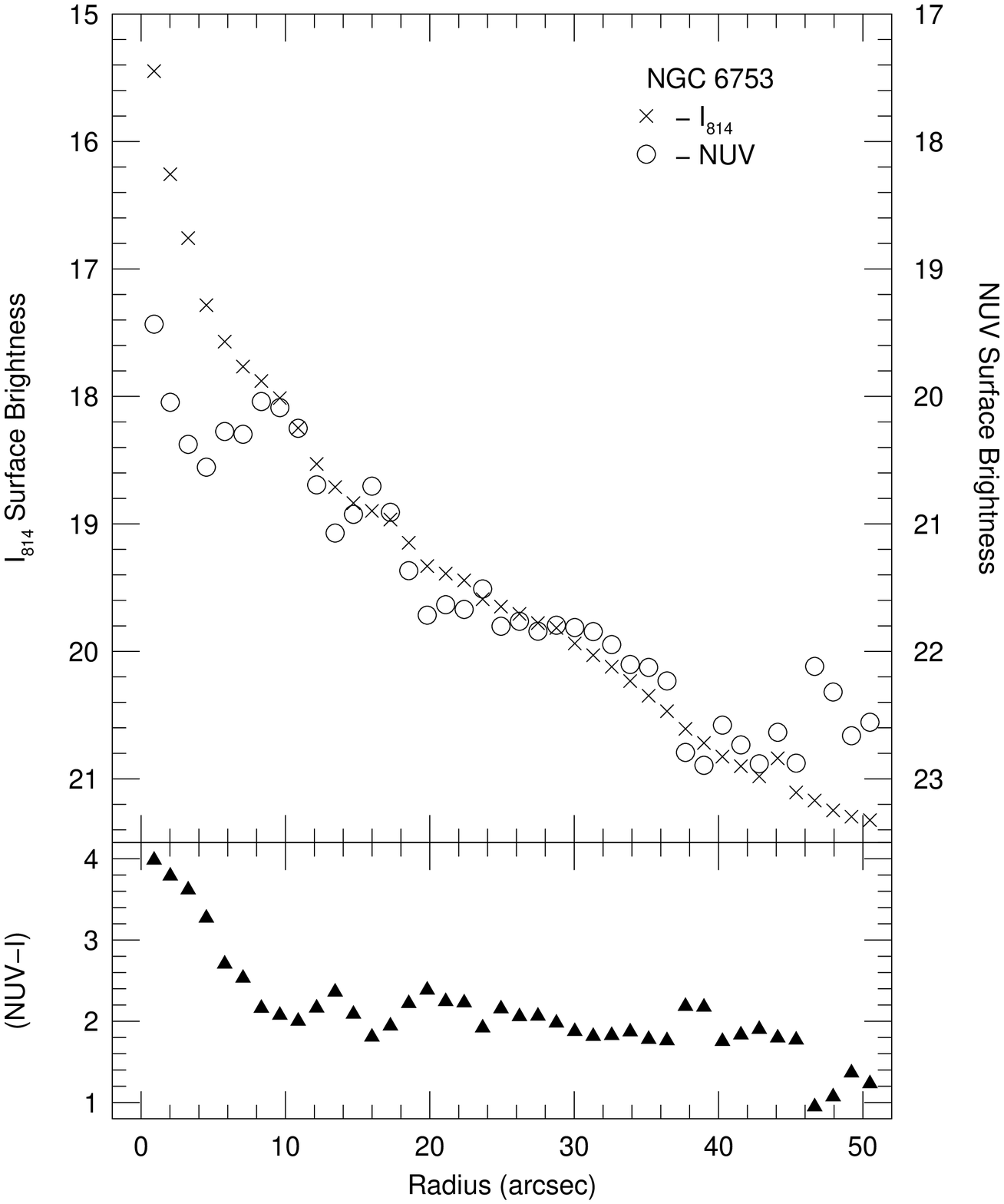}
\newpage
\plotone{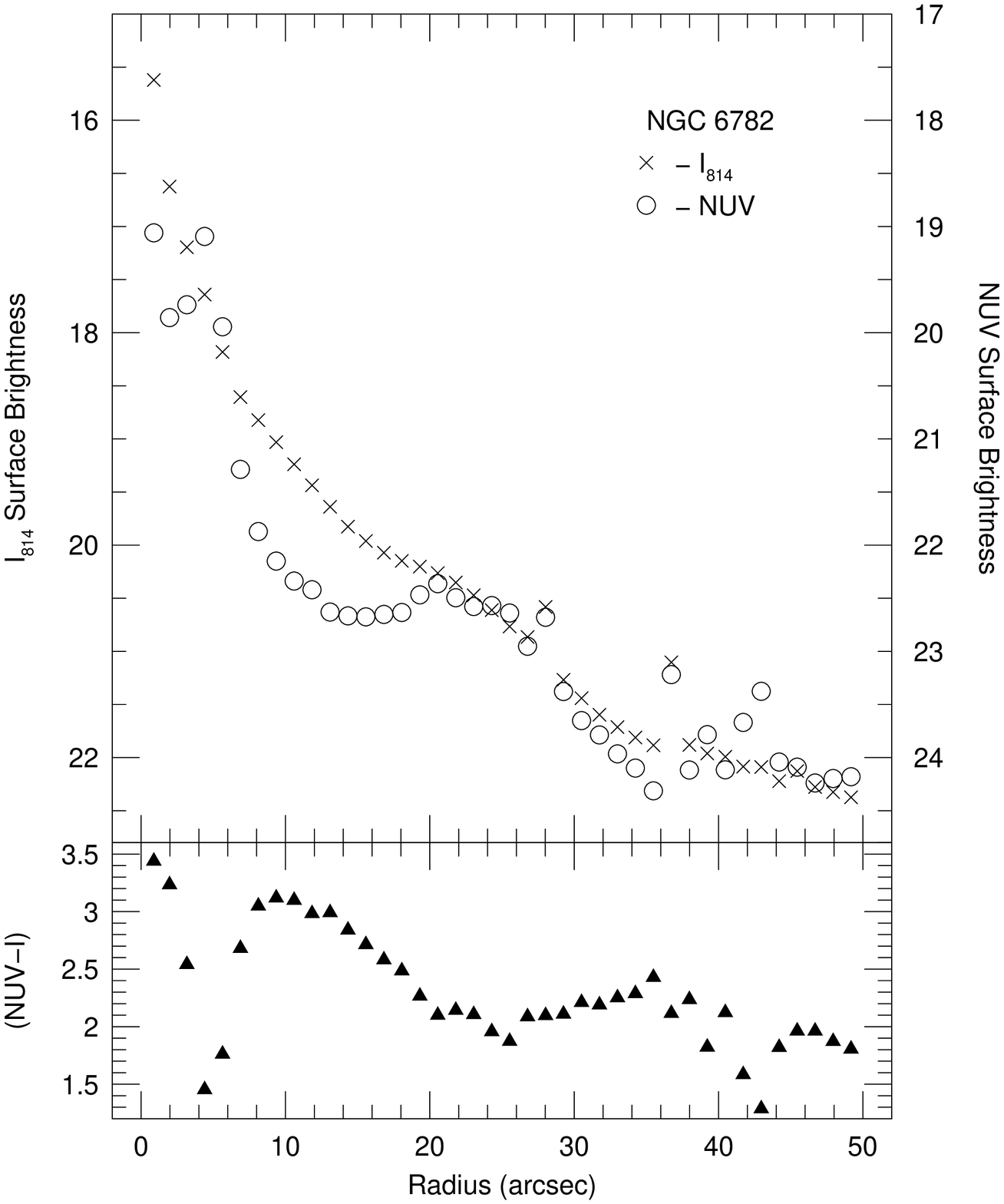}
\newpage
\plotone{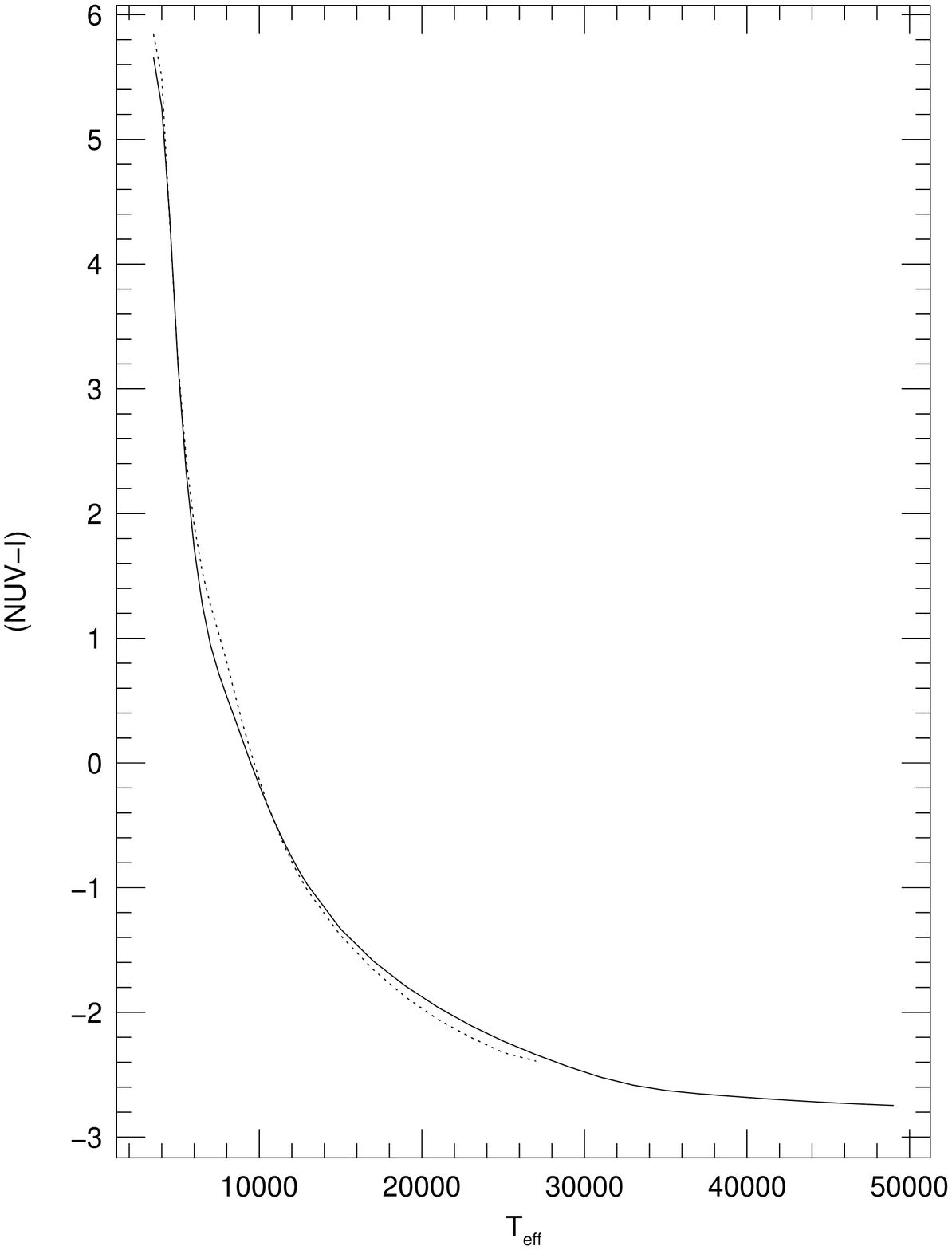}
\newpage
\plotone{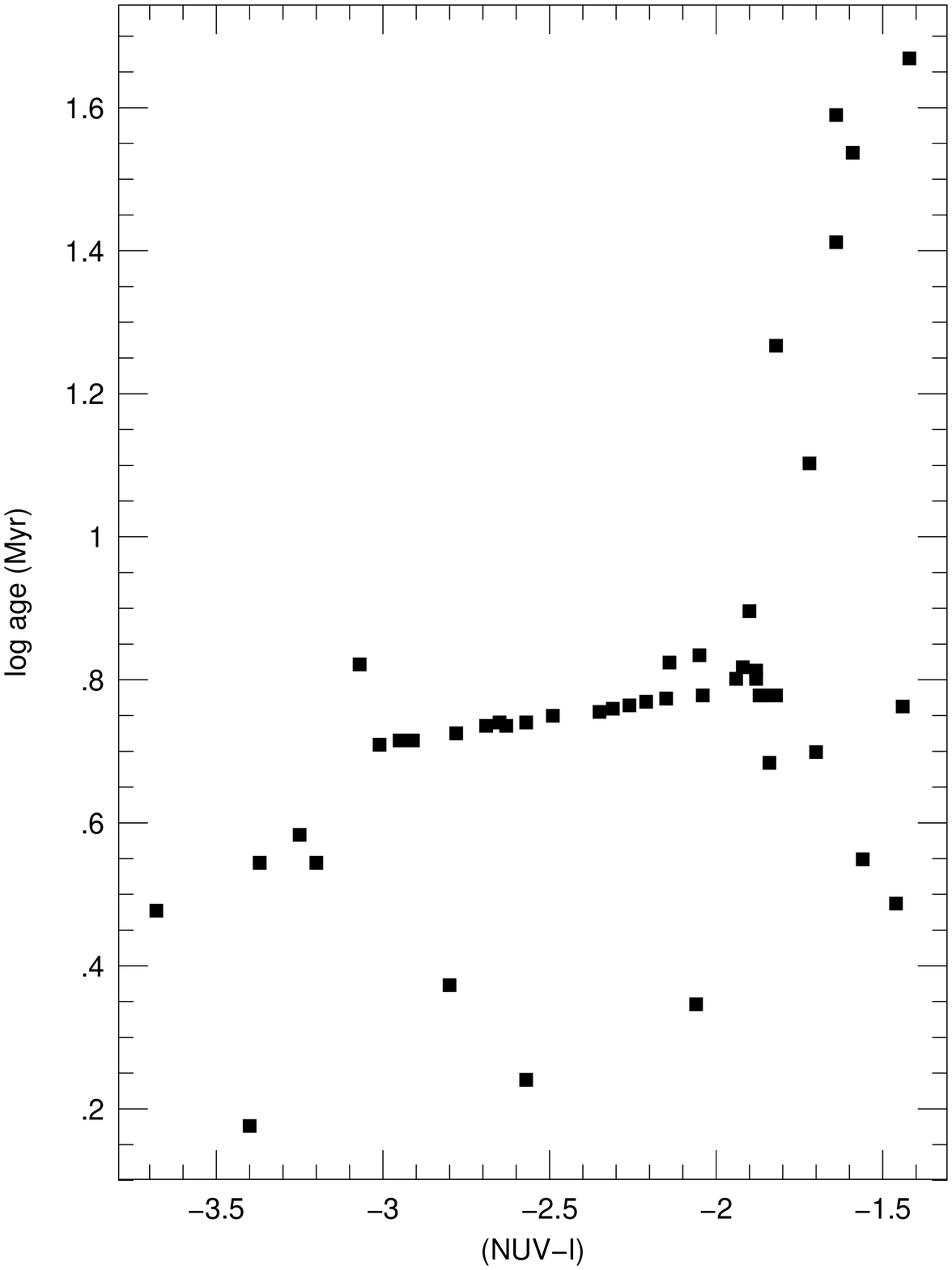}


\clearpage

\begin{references}
\reference{aea} Abraham, R.G., Ellis, R.S., Fabian, A.C., Tanvir, N.R. \& 
Glazebrook, K. 1999, \mnras, 308, 569
\reference{bba} Baggett, W.E., Baggett, S.M. \& Anderson, K.S.J. 1998, \aj, 116,
1626
\reference{bfp} Baldwin, J.R., Frogel, J.A. \& Persson, S.E. 1973, \apj, 184,
427
\reference{wf2} Biretta, J.A., et al. 2000, WFPC2 Instrument Handbook, Version 
5.0 (Baltimore: STScI)
\reference{blo} Block, D.L., Bertin, G., Stockton, A., Grosb\o l, P., Moorwood,
A.F.M. \& Peletier, R.F. 1994, \aap, 288, 365
\reference{b86} Bothun, G.D. 1986, \aj, 91, 507
\reference{fst} Bowyer, S., Sasseen, T.P., Wu, X. \& Lampton, M. 1995, \apjs, 
96, 461
\reference{b93} Bruzual, A.G. \& Charlot, S. 1993, \apj, 405, 538
\reference{bns} Bushouse, H. \& Simon, B. 1998, Synphot User's Guide 
(Baltimore: STScI)
\reference{bnc} Buta, R. \& Crocker, D.A. 1993, \aj, 105, 1344
\reference{csr} Buta, R. 1995, \apjs, 96, 39
\reference{bea} Byrd, G., Rautiaine, P., Salo, H., Buta, R. \& Crocker, D.A. 
1994, \aj, 108, 476
\reference{rip} Cardelli, J.A., Clayton, G.C. \& Mathis, J.S. 1989, \apj, 345, 
245
\reference{cdt} Castellanos, M., D\'{\i}az, A.I. \& Terlevich, E. 2002, \mnras, 
329, 315
\reference{cea} Crocker, D.A., Baugus, P.D. \& Buta, R. 1996, \apjs, 105, 353
\reference{deG} de Grijs, R. 1998, \mnras, 299, 595
\reference{deJ} de Jong, R.S. 1996, \aap, 313, 377
\reference{rc3} de Vaucouleurs, G., de Vaucouleurs A., Corwin, H.G., Jr., Buta,
R.J., Paturel, G. \& Fouqu\'e, P. 1991, Third Reference Catalogue of Bright
Galaxies (Springer-Verlag:  New York) (RC3)
\reference{bap} Eskridge, P.B., Frogel, J.A., Pogge, R.W., Quillen, A.C., 
Berlind, A.A., Davies, R.L., DePoy, D.L., Gilbert, K.M., Houdashelt, M.L., 
Kuchinski, L.E., Ram\'{\i}rez, S.V., Sellgren, K., Stutz, A., Terndrup, D.M. \&
Tiede, G.P. 2002, \apjs, 143, 73
\reference{f85} Frogel, J.A. 1985, \apj, 298, 528
\reference{fqp} Frogel, J.A., Quillen, A.C., \& Pogge, R.W. 1996, in New 
Extragalactic Perspectives in the New South Africa, ed.~D.~Block \& 
J.M.~Greenberg, (Dordrecht: Kluwer), 65
\reference{g89} Garnett, D.R. 1989, \apj, 345, 282
\reference{hs} Hackwell, J.A. \& Schweizer, F. 1983, \apj, 265, 643
\reference{hea} Harris, J., Calzetti, D., Gallagher, J.S., III, Conselice, C.J.
\& Smith, D.A. 2001, \aj, 122, 3046
\reference{zps} Holtzman, J.A., Burrows, C.J., Casertano, S., Hester, J.J.,
Trauger, J.T., Watson, A.M. \& Worthey, G. 1995, \pasp, 107, 1065
\reference{k85} Kent, S.M., 1985, \apjs, 59, 115
\reference{kea} Kinney, A.L., Calzetti, D., Bohlin, R.C., McQuade, K., 
Storchi-Bergmann, T. \& Schmitt, H.R. 1996, \aj, 114, 592
\reference{k77} Kormendy, J., 1977, \apj, 217, 406
\reference{kaw} Kron, R.G., Annis, J. \& Wilhite, B.C. 2000 \baas, 31, 9.15
\reference{ku0} Kuchinski, L.E., Freedman, W.L., Madore, B.F., Trewhella, M., 
Bohlin, R.C., Cornett, R.H., Fanelli, M.N., Marcum, P.M., Neff, S.G., 
O'Connell, R.W., Roberts, M.S., Smith, A.M., Stecher, T.P. \& Waller, W.H. 
2000, \apjs, 131, 441
\reference{ku1} Kuchinski, L.E., Madore, B.F., Freedman, W.L., \& Trewhella, M. 
2001, \aj, 122, 729
\reference{m88} Majewski, S.R. 1988, in Toward Understanding Galaxies at Large
Redshift, ed. R.G.~Kron \& A.~Renzini, (Dordrecht: Kluwer), 127
\reference{mea} Marcum, P.M., O'Connell, R.W., Fanelli, M.N., Cornett, R.H., 
Waller, W.H., Bohlin, R.C., Neff, S.G., Roberts, M.S., Smith, A.M., Cheng, 
K.-P., Collins, N.R., Hennessy, G.S., Hill, J.K., Hill, R.S., Hintzen, P.,
Landsman, W.B., Ohl, R.G., Parise, R.A., Smith, E.P., Freedman, W.L., 
Kuchinski, L.E., Madore, B.F., Angione, R., Palma, C., Talbert, F. \& Stecher, 
T.P. 2001, \apjs, 132, 129
\reference{pea} Peng, C.Y., Ho, L.C., Impey, C.D. \& Rix, H.-W. 2002, \aj, 124,
266
\reference{pfc} Persson, S.E., Aaronson, M., Cohen, J.G., Frogel, J.A. \& 
Matthews, K. 1983, \apj, 266, 105
\reference{r98} Rhoads, J.E. 1998, \aj, 115, 472
\reference{rea} Romaniello, M., Panagia, N., Scuderi, S. \& Kirshner, R.P. 
2002, \aj, 123, 915
\reference{sns} Sage, L.J. \& Solomon, P.M. 1989, \apj, 342, L15
\reference{cag} Sandage, A. \& Bedke, J. 1994, The Carnegie Atlas of Galaxies
(Carnegie Institute of Washington:  Washington DC)
\reference{sch} Schaller, G., Schaerer, D., Meynet, G. \& Maeder, A. 1992, 
\aaps, 96, 269
\reference{sfd} Schlegel, D.J., Finkbeiner, D.P. \& Davis, M. 1998, \apj, 500,
525
\reference{wnk} Walterbos, R.A.M. \& Kennicutt, R.C., 1987, \aaps, 69, 311
\reference{pzp} Windhorst, R.A., Taylor, V.A., Jansen, R.A., Odewahn, S.C., 
Chiarenza, C.A.T., Conselice, C.J., de Grijs, R., de Jong, R.S., MacKenty, J.,
Eskridge, P.B., Frogel, J.A., Gallagher, J.S., III, Hibbard, J.E., Matthews, 
L.D. \& O'Connell, R.W. 2002, \apjs, 143, 113
\reference{yts} Yahil, A., Tammann, G.A. \& Sandage, A. 1977, \apj, 217, 903
\reference{fzw} Zwicky, F. 1955, \pasp, 67, 232
\end{references}
\end{document}